\documentclass[preprint,12pt,3p]{elsarticle}

\usepackage{graphicx}
\usepackage{amssymb}
\usepackage{amsthm}
\usepackage{amsmath}
\usepackage{hyperref}

\DeclareMathOperator*{\argmax}{arg\,max}

\begin{document}

\begin{frontmatter}

\title{A new look at calendar anomalies: Multifractality and day of the week effect}
%\title{Calendar anomalies from a multifractal perspective}

\author[addr1]{Darko Stosic\corref{cor1}}
\address[addr1]{Centro de Inform\'atica, Universidade Federal de Pernambuco, Av. Luiz Freire s/n, 50670-901, Recife, PE, Brazil}
\cortext[cor1]{Corresponding author.}
\ead{ddstosic@bu.edu}

\author[addr1]{Dusan Stosic}
\ead{dbstosic@bu.edu}

\author[addr2]{Irena Vodenska}
\address[addr2]{Department of Administrative Sciences, Metropolitan College, Boston University, 1010 Commonwealth Avenue, Boston, MA 02215, USA}
\ead{vodenska@bu.edu}

\author[addr3]{H. Eugene Stanley}
\address[addr3]{Center for Polymer Studies and Department of Physics, Boston University, 590 Commonwealth Avenue, Boston, MA 02215, USA}
\ead{hes@bu.edu}

\author[addr4]{Tatijana Stosic}
\address[addr4]{Departamento de Estat\'{i}stica e Inform\'{a}tica, Universidade Federal Rural de Pernambuco, Rua Dom Manoel de Medeiros s/n, Dois Irm\~{a}os, 52171-900, Recife, PE, Brazil}
\ead{tastosic@gmail.com}

\begin{abstract}
Stock markets can become inefficient due to calendar anomalies known as day-of-the-week effect. Calendar anomalies are well-known in financial literature, but the phenomena remain to be explored in econophysics. In this paper we use multifractal analysis to evaluate if the temporal dynamics of market returns also exhibits calendar anomalies such as day-of-the-week effects. We apply the multifractal detrended fluctuation analysis (MF-DFA) to daily returns of market indices around the world for each day of the week. Our results indicate that individual days of the week are characterized by distinct multifractal properties. Monday returns tend to exhibit more persistent behavior and richer multifractal structures than other day-resolved returns. Shuffling the series reveals that multifractality arises both from a broad probability density function and from long-term correlations. From the time-dependent multifractal analysis we find that multifractal spectra for Monday returns are much wider than for other days during periods of financial crises. The presence of day-of-the-week effects in multifractal dynamics of market returns motivates further research on calendar anomalies from an econophysics perspective.
\end{abstract}

\end{frontmatter}

% \linenumbers

% on ripple: http://www.bbc.com/news/technology-42541390
% on dogecoin: http://www.bbc.com/news/business-42602038

\section{Introduction}\label{secintro}
Market prices should incorporate and reflect all available information, at any point in time, according to the Efficient Market Hypothesis (EMH)~\cite{ref1,ref2}. Yet various studies~\cite{ref3,ref4,ref5,ref6} show that financial markets often become inefficient and their behavior no longer follows that of a random walk. Stock markets can instead deviate from the rules of EMH in the form of anomalies. Anomalies can be broadly categorized into calendar, fundamental and technical anomalies~\cite{ref7}. The most studied set of pricing anomalies are calendar or seasonal anomalies that represent systematic patterns of security returns around certain calendar points. Calendar anomalies include day-of-the-week effect~\cite{ref8}, turn-of-the-month~\cite{ref9}, turn-of-the-year~\cite{ref10} and holiday effect~\cite{ref11}. The day-of-the-week effect refers to the tendency of stocks to exhibit significantly higher returns on one particular day compared to other days in the week. Cross~\cite{ref12} first proved evidence of day-of-the-week effects on the Standard and Poor's index -- it was found that price returns are significantly negative on Monday. Since then, this phenomenon has been extensively studied and discovered in other financial markets such as stock markets~\cite{ref13,ref14,ref15}, exchange rates~\cite{ref16,ref17}, bonds~\cite{ref18}, crude oil~\cite{ref19}, gold~\cite{ref20} and cryptocurrencies~\cite{ref21}. For a detailed review of seasonal anomalies the reader can look at Refs.~\cite{ref22} and~\cite{ref23}.

Financial markets have also attracted much attention from researchers in related fields such as econophysics. Econophysics has paved the road for new perspectives and understanding of financial markets by drawing concepts from statistical physics such as fractals and multifractals~\cite{ref24,ref25,ref26,ref27}, information theory~\cite{ref28,ref29} and network structures~\cite{ref30,ref31,ref32} (see Ref.~\cite{refx1} and references therein for a comprehensive review). While many well-known conclusions in literature on an array of financial markets (including market indices, stocks, exchange rates and commodities) can be attributed to econophysics, there are still a number of important phenomena to be investigated from this perspective. To the best of our knowledge, one such phenomena that remains to be unearthed is the calendar anomaly, and our study is designed as a contribution in this direction.

In this paper we use multifractal analysis to evaluate if temporal dynamics of market returns exhibit calendar anomalies such as day-of-the-week effects. We apply Multifractal Detrended Fluctuation Analysis (MDFA)~\cite{ref34} to daily returns of market indices around the world for each day of the week (Monday returns, Tuesday returns and so on). We then compare properties of the resulting multifractal spectra that reveal distinct behavior in the underlying stochastic process. Economic literature states that for some markets this effect disappears as the market become more efficient~\cite{ref35,ref36}. To observe this behavior over time, we perform time-dependent multifractal analysis on the United States (GSPC) market by calculating multifractal spectra of the return series in a sliding window. This novel approach permits us to analyze the temporal evolution of multifractal parameters which are related to different properties of market fluctuation, leading to better understanding of the underlying stochastic processes. The paper is organized as follows: Section~\ref{secmethod} introduces MF-DFA and the time-dependent method. Section~\ref{secdata} describes the market data. Section~\ref{secres} presents the results and discussions. Section~\ref{secconcl} draws the conclusions.

\section{Methods}\label{secmethod}
While fractal processes are characterized by long-term correlations that are described by a single scaling exponent, in multifractal time series subsets with small and large fluctuations can scale differently, and the analysis of long-term correlations results in a hierarchy of scaling exponents~\cite{ref34}. Multifractal analysis of temporal series can be performed using different methods, such as wavelet transform modulus maxima (WTMM) method~\cite{ref37}, multifractal detrended fluctuation analysis (MF-DFA) method~\cite{ref34} and multifractal detrending moving average method (MF-DMA)~\cite{ref38}. In this work we employ MF-DFA that has been found to produce reliable results~\cite{ref39}, and has been widely used to analyze physiological signals~\cite{ref40,ref41,ref46}, geophysical data~\cite{ref42}, weather data~\cite{ref43}, and financial time series~\cite{ref45}.

The implementation of the MF-DFA algorithm can be described as follows~\cite{ref34}.

\begin{enumerate}[i]
\item The first step is the integration of the original series $x(i),i=1,\dots,N$ to produce
\begin{align}
X(k) = \sum_{i=1}^k[x(i)-\langle x\rangle],\quad k=1,\dots,N,
\end{align}
where $\langle x\rangle =\frac{1}{N}\sum_{i=1}^k x(i)$ is the average.
\item Next the integrated series $X(k)$ is divided into $N_n=int(N/n)$ non-overlapping segments of length $n$ and in each segment $\nu=1,\dots,N_n$ the local trend $X_{n,\nu} (k)$ is estimated as a linear or higher order polynomial least square fit and subtracted from $X(k)$.
\item The detrended variance
\begin{align}
F^2(n,\nu) = \frac{1}{n}\sum_{k=(\nu-1)n+1}^{\nu n}\left[X(k)-X_{n,\nu}(k)\right]^2
\end{align}
is calculated for each segment and then averaged over all segments to obtain the $q$th order fluctuation function
\begin{align}
F_q(n) = \left\{\frac{1}{N_n}\sum_{\nu=1}^{N_n}[F^2(n,\nu)]^{q/2}\right\}^{1/q},
\end{align}
where, in general, $q$ can take on any real value except zero.
\item 	Repeating this calculation for all box sizes provides the relationship between the fluctuation function $F_q(n)$ and box size $n$. $F_q(n)$ increases with $n$ according to a power law $F_q (n)\sim n^{h(q)}$ if long-term correlations are present. The scaling exponent $h(q)$ is obtained as the slope of the linear regression of $\log F_q(n)$ versus $\log n$.
\end{enumerate}

The power law exponent, $h(q)$, is called the generalized Hurst exponent, where for stationary time series $h(2)$ is identical to the well-known Hurst exponent $H$. For positive $q$ values $h(q)$ describes the scaling behavior of large fluctuations, while for negative $q$ values $h(q)$ describes the scaling behavior of small fluctuations. $h(q)$ is independent of $q$ for monofractal time series and a decreasing function of $q$ for multifractal time series.

Generalized Hurst exponents are related to the Renyi exponents $\tau(q)$ defined by the standard partition function-based multifractal formalism $\tau(q)=qh(q)-1$. For monofractal signals $\tau(q)$ is a linear function of q (as $h(q)=const$.) and for multifractal signals $\tau(q)$ is a nonlinear function of $q$. A multifractal process can also be characterized by the singularity spectrum $f(\alpha)$ which is related to $\tau(q)$ through the Legendre transform:
\begin{align}
\alpha(q) = \frac{d\tau(q)}{dq},
\end{align}
\begin{align}
f(\alpha(q)) = q\alpha(q) - \tau(q),
\end{align}
where $f(\alpha)$ is the fractal dimension of the support of singularities in the measure with Lipschitz-Holder exponent $\alpha$. The singularity spectrum of monofractal signal is represented by a single point in the $f(\alpha)$ plane, whereas multifractal process yields a single humped function.

Multifractal spectra reflect the level of complexity of the underlying stochastic process, and can be characterized by a set of three parameters which are determined as follows. The singularity spectra are fitted to a fourth degree polynomial:
\begin{align}
f(\alpha)=A+B(\alpha-\alpha_0 )+C(\alpha-\alpha_0 )^2+D(\alpha-\alpha_0 )^3+E(\alpha-\alpha_0 )^4
\end{align}
and the multifractal spectrum parameters are found as the position of maximum $\alpha_0=\argmax_\alpha f(\alpha)$, width of the spectrum $W=\alpha_{max}-\alpha_{min}$ obtained from extrapolating the fitted curve to zero, and skew parameter $r=(\alpha_{max}-\alpha_0)/(\alpha_0-\alpha_{min})$ where $r=1$ for symmetric shapes, $r>1$ for right-skewed shapes and $r<1$ for left-skewed shapes. These three parameters can be used to evaluate the complexity of the underlying process. A small value of $\alpha_0$ means that the process is correlated and more regular in appearance. The width $W$ of the spectrum measures the degree of multifractality of the process -- wider range of fractal exponents leads to ``richer'' structures. The skew parameter $r$ indicates which fractal exponents are dominant: $f(\alpha)$ spectrum is right-skewed ($r>1$) and the process is characterized by ``fine structure'' (small fluctuations) if high fractal exponents are dominant, whereas the process is more regular or smooth, $f(\alpha)$ spectrum is left-skewed ($r<1$) and fractal exponents describe the scaling of large fluctuations if low fractal exponents are dominant. In summary, a signal with a high value of $\alpha_0$, a wide range $W$ of fractal exponents (higher degree of multifractality), and a right-skewed shape ($r>1$) may be considered more complex than one with the opposite characteristics~\cite{ref46}.

The two sources of multifratality in a time series are: (i) broad probability density function for the values of the time series, and (ii) different long-term correlations for small and large fluctuations. The type of multifractal can be found by randomly shuffling the series and analyzing its behavior. For multifractals of type (ii) the shuffled series exhibits simple random behavior (since long-term correlations are destroyed) and the width of the $f(\alpha)$ spectrum is reduced to a single point, for multifractals of type (i) the width of the $f(\alpha)$ spectrum remains the same (since probability density cannot be removed), and for multifractals of type (i) and (ii) the shuffled series shows weaker multifractality than the original series~\cite{ref34}.

The time-dependent MF-DFA algorithm is based on the sliding window technique and yields a temporal evolution of multifractality in the system. Given a time series $x=x_1,\dots,x_N$, many sliding windows $z_t=x_{1+t\Delta},\dots,x_{w+t\Delta}, t=0,1,\dots,\left[\frac{N-w}{\Delta}\right]$ are constructed, where $w\leq N$ is the window size, $\Delta\leq w$ is the sliding step, and operator $[.]$ denotes taking integer part of the argument. The values of the time series in each window $z_t$ are then used to calculate the multifractal spectrum at a given time $t$ using the method described above. This allows us to obtain time evolutions for the three complexity parameters.

\begin{figure}[!htb]
\begin{center}
\includegraphics[width=1\textwidth]{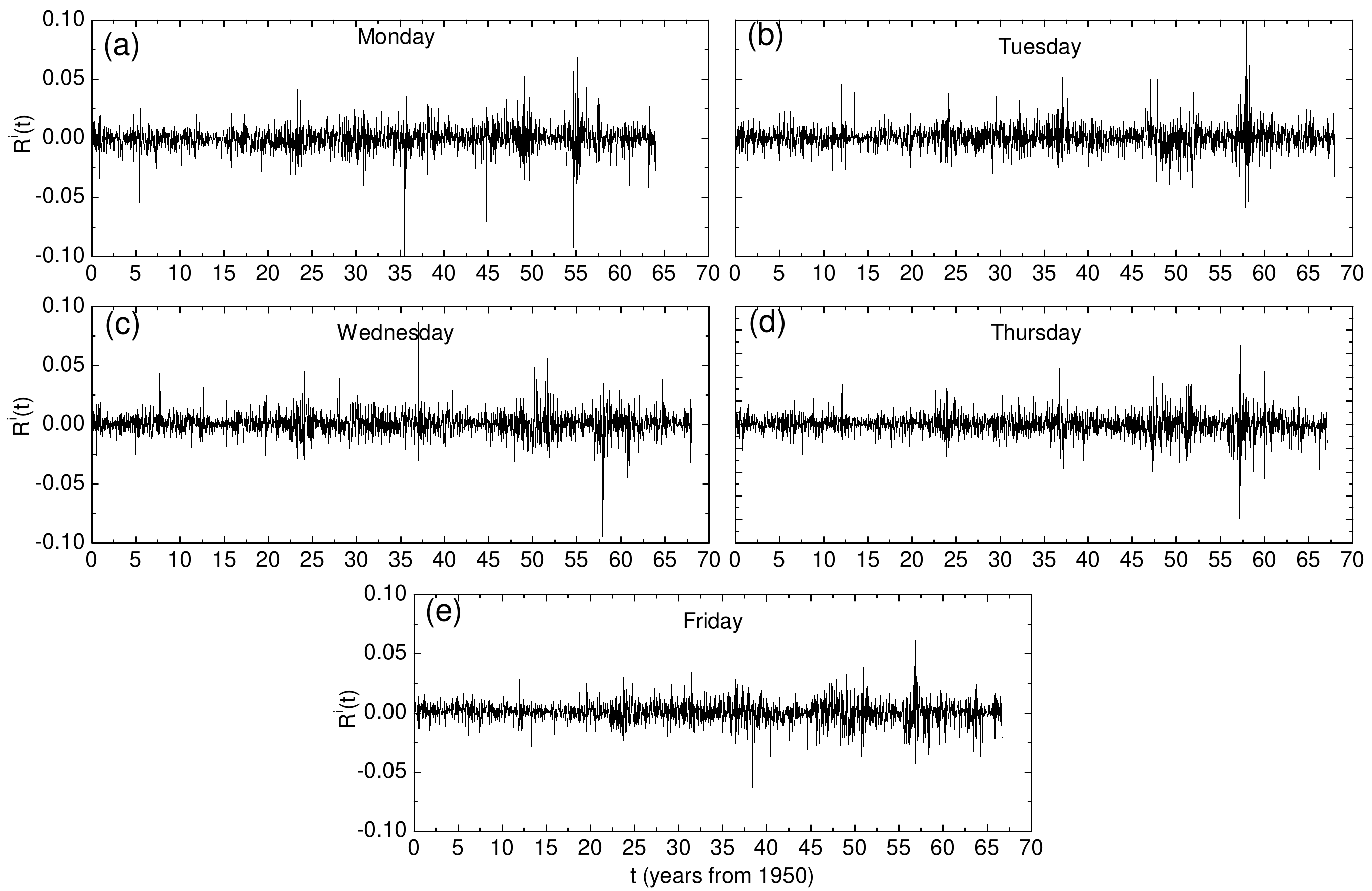}
\caption{Time series for (a) Monday, (b) Tuesday, (c) Wednesday, (d) Thursday and (e) Friday day-resolved price returns $R^i$ of the United States (GSPC) market index.
\label{figdata}}
\end{center}
\end{figure}

\section{Data}\label{secdata}
We analyze the time series of $19$ major stock market indices that appear in the website~\url{https://finance.yahoo.com/world-indices/} as listed in Table~\ref{tabdata}. The period under study spans the earliest recorded date to 2019. For each of the market indices we calculate the daily logarithmic returns in closing price $S(t)$:
\begin{align}
R_t\equiv\ln\frac{S(t+1)}{S(t)}
\end{align}
and construct time series from the returns $R_t$ for each day of the week (Monday returns, Tuesday returns and so on):
\begin{align}
R^i = \{R_{i}, R_{i+5}, \dots, R_{i+5\left[\frac{N}{5}\right]}\},
\end{align}
where $i$ denotes the index of the day, and operator $[.]$ denotes taking integer part of the argument.
Fig.~\ref{figdata} reveals that fluctuations in the returns vary between different days. While Monday exhibits the most pronounced negative returns, fluctuations for other days dominate at specific time intervals. This is a well known day-of-the-week effect which was found for US market \cite{ref8, ref13}.

The MF-DFA method is applied to day-resolved returns $R^i$ of major stock market indices, where local trends are fitted with a second degree polynomial $m=2$. Next we perform a fourth order polynomial regression on the singularity spectra $f(\alpha)$ to determine the position of maximum $\alpha_0$ and the zeros of the polynomial, $\alpha_{\text{max}}$ and $\alpha_{\text{min}}$. From the polynomial fits we calculate three measures of complexity: the position of maximum $\alpha_0$, the width of the spectrum $W=\alpha_{\text{max}}-\alpha_{\text{min}}$, and the skew parameter $r=(\alpha_{\text{max}}-\alpha_0)/(\alpha_0-\alpha_{\text{min}})$. These parameters are then used to determine the multifractal behavior of day-resolved price returns.

\begin{table}[!t]
\caption{Information on analyzed time series for major market indices.}
\label{tabdata}
\centering
\scalebox{0.75}{
\begin{tabular}{lllr}
\hline
Market	&	Country	&	Index	&	Period	\\
\hline
All Ordinares	&	Australia	&	AORD	&	8/3/1984 - 12/26/2018	\\
S\&P500/ASX 200	&	Australia	&	AXJO	&	11/22/1992 - 12/26/2018	\\
BEL 20	&	Belgium &	BFX	&	4/9/1991 - 12/24/2018	\\
IBOVESPA	&	Brazil &	BVSP	&	4/27/1993 - 12/21/2018	\\
Dow30	&	United States &	DJI	&	1/29/1985	-	12/26/2018	\\
CAC 40	&	France	&	FCHI &	3/1/1990 - 12/24/2018	\\
DAX Performance	&	Germany	&	GDAXI	&	12/30/1987 - 12/27/2018	\\
S\&P500	&	United States	&	GSPC &	1/3/1950 - 12/24/2018	\\
S\&P/TSX Composite	&	Canada &	GSPTSE	&	6/29/1979 - 12/24/2018	\\
Hang Seng Index	&	Hong Kong &	HIS	&	12/31/1986 - 12/27/2018	\\
IPSA Santiago de Chile	&	Chile	&	IPSA	&	1/2/2002 - 12/26/2018	\\
Nasdaq	&	United States	&	IXIC	&	2/5/1971 - 12/26/2018	\\
Jakarta Composite	&	Indonesia	&	JKSE	&	7/1/1997 - 12/27/2018	\\
KOSPI Composite	&	South Korea &	KS11	&	7/1/1997 - 12/26/2018	\\
Merval	&	Argentina	&	MERV	&	10/8/1996 - 12/26/2018	\\
IPC Mexico	&	Mexico	&	MXX	&	11/8/1991 - 12/26/2018	\\
Nikkei 225	&	Japan	&	N225	&	1/5/1965 - 12/27/2018	\\
NYSE Composite	&	United States	&	NYA	&	12/31/1965 - 12/26/2018	\\
TSEC weighted	&	Taiwan	&	TWII	&	7/2/1997 - 12/27/2018	\\
\hline
\end{tabular}
}
\end{table}

\section{Results}\label{secres}
\subsection{Day-of-the-week effect}
\begin{figure}[!t]
\begin{center}
\includegraphics[width=0.7\textwidth]{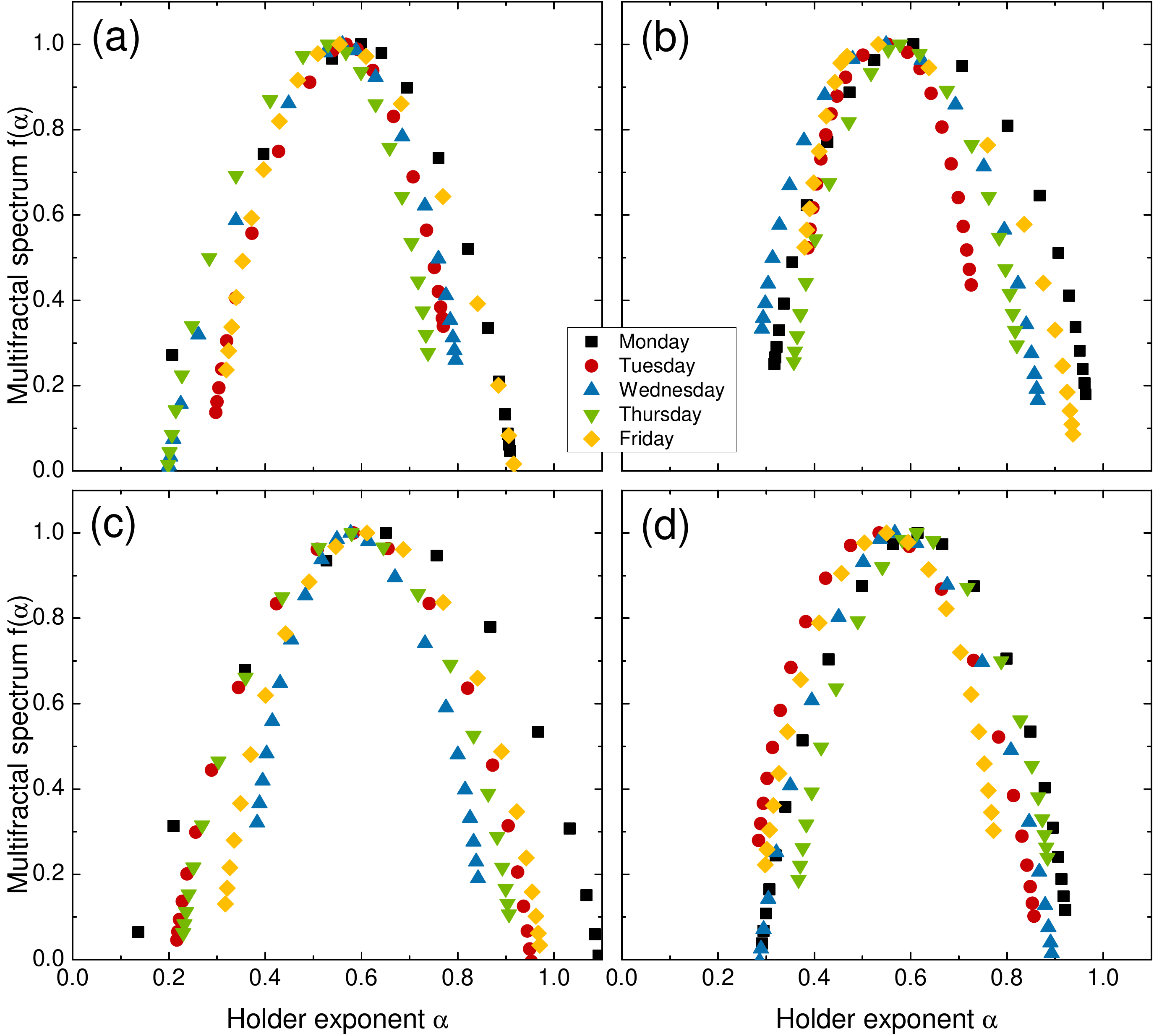}
\caption{Multifractal spectrum $f(\alpha)$ for day-resolved price returns $R^i$ of (a) United States (GSPC), (b) South Korea (KS11), (c) Chile (IPSA) and (d) France (FCHI) market indices.
\label{figspectrafour}}
\end{center}
\end{figure}

Complexity measures derived from the singularity spectra are used to study multifractal behavior of price returns for every day of the week. We first consider multifractality in day-resolved price returns from four distinct markets: United States (GSPC), South Korea (KS11), Chile (IPSA) and France (FCHI). The multifractal spectra for each day using the four markets are illustrated in Fig.~\ref{figspectrafour}. We observe that day-of-the-week effects lead to significant differences in multifractal behavior: (i) positions of maxima $\alpha_0$ are shifted to the right $(\alpha_0>0.5)$ for Monday returns and (ii) spectrum widths $W$ are wider on Monday than on returns from other days. There seem to be no consistent differences in the skew parameter $r$, which implies that both large and small fluctuations are present for different days of the week (e.g., see Tab.~\ref{tabcomplexity}). These results indicate that Monday returns exhibit more persistent behavior and richer multifractal structures, which leads to more complex time series than other day returns. Our findings are consistent with results obtained from Ref.~\cite{ref13}, which indicated that Monday has the largest anomalies (day-of-the-week effect) because of the weekend gap in trading hours. Other days of the week do not exhibit any visible patterns in multifractal behavior for either the position or width of the spectrum.

\begin{figure}[!b]
\begin{center}
\includegraphics[width=1\textwidth]{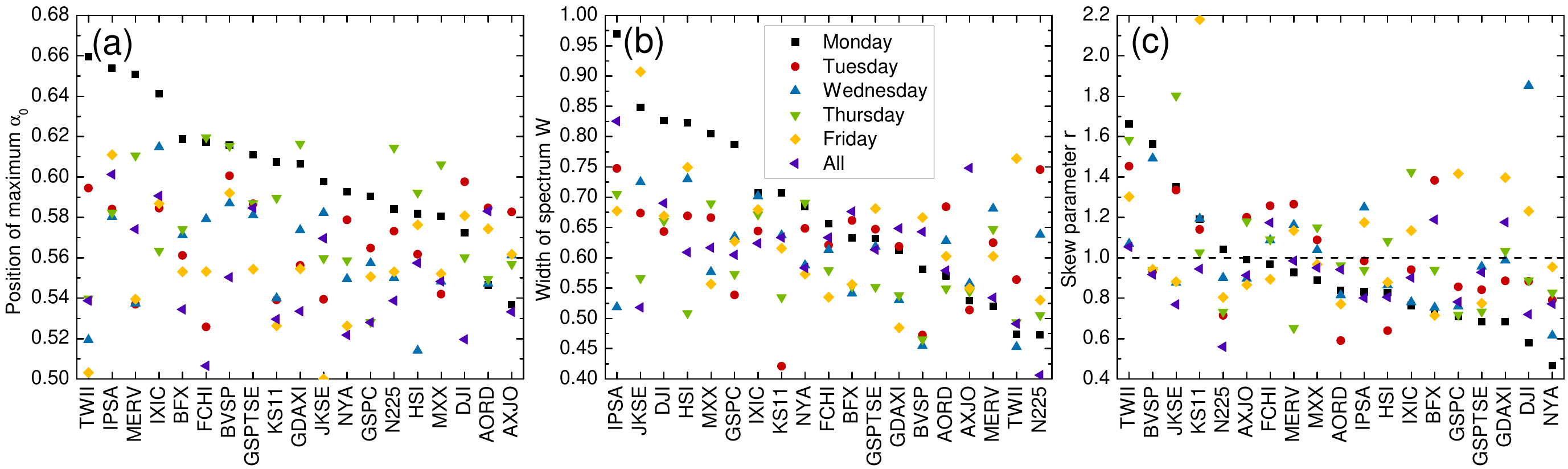}
\caption{Complexity parameters for day-resolved price returns of the market indices listed in Tab.~\ref{tabdata}, sorted from largest to smallest. 
\label{figcomplexity}}
\end{center}
\end{figure}

We expand our investigation to other markets listed in Tab.~\ref{tabdata}. Fig.~\ref{figcomplexity} reveals that multifractal spectra of Monday returns are dominantly right-shifted $(\alpha_0>0.5)$ compared to other days for most analyzed markets. 
Notable exceptions include Unated States (DJI), Australia (AORD,AXJO) where Tuesday returns are more persistent, and Japan (N225) where Thursday returns exhibit stronger persistency.
The width of the multifractal spectrum displays similar tendencies as its position, where Monday returns possess broader multifractal widths. Yet we find that more markets tend to have other days with richer multifractal structures: multifractal spectral are the widest for Friday returns in Taiwan (TWII) and for Tuesday returns in Japan (N225) and Australia (AORD), as opposed to markets with dominant Monday returns considered so far. 
It has been noted that day-of-the-week effect occurs on different distinct days of the week for different markets \cite{ref13}. 
Considering both parameters  $\alpha_0$ and $W$, we observe that  North American, European,  some Asian (South Corea, Indonesia and Honkong)  and some Latino American markets  (Chile and Mexico) tend to show both stronger persistency and stronger multifractality for Monday returns, while  for Australia, Indonesia and Taiwan this tendency  is found  for Tuesday returns. The  is  also in agreement with literature, where was found that some  Asian markets  display Tusday anomaly, one day out of phase with North American markets due to different time zones \cite{refx3}. Patterns in the skew of multifractal spectra for a given day of the week are again hard to discern across distinct markets, where both small and large fluctuations exist. Values of the multifractal complexity parameter are listed in Tab.~\ref{tabcomplexity}. Our results indicate that while most markets exhibit more complex behavior for Monday returns, some markets have other days with largest anomalies (day-of-the-week effect) such as Tuesday, Thursday and Friday returns. This is expected from literature where it was found that different day-of-the-week effects exist for different markets~\cite{ref13}.

\begin{table}[!htb]
\caption{Multifractal parameters $\alpha_0$, $W$, and $r$ for day-resolved price returns $R^i$ of major market indices}
\label{tabcomplexity}
\centering
\scalebox{0.6}{
\begin{tabular}{lcccccccccccccccccc}
\hline
\multicolumn{1}{l}{} & \multicolumn{3}{l}{Monday} & \multicolumn{3}{l}{Tuesday} & \multicolumn{3}{l}{Wednesday} & \multicolumn{3}{l}{Thursday} & \multicolumn{3}{l}{Friday} & \multicolumn{3}{l}{All} \\
\hline
Market & $\alpha_0$ & $W$ & $r$ & $\alpha_0$ & $W$ & $r$ & $\alpha_0$ & $W$ & $r$ & $\alpha_0$ & $W$ & $r$ & $\alpha_0$ & $W$ & $r$ & $\alpha_0$ & $W$ & $r$ \\
\hline
AORD		&	0.547	&	0.570	&	0.837	&	0.585	&	0.684	&	0.590	&	0.547	&	0.628	&	0.815	&	0.549	&	0.549	&	0.963	&	0.574	&	0.603	&	0.771	&	0.583	&	0.579	&	0.942	\\
AXJO		&	0.537	&	0.529	&	0.990	&	0.583	&	0.514	&	1.201	&	0.561	&	0.558	&	0.897	&	0.557	&	0.544	&	1.180	&	0.562	&	0.550	&	0.866	&	0.533	&	0.748	&	0.913	\\
BFX			&	0.619	&	0.633	&	0.730	&	0.561	&	0.662	&	1.383	&	0.571	&	0.541	&	0.754	&	0.574	&	0.553	&	0.940	&	0.553	&	0.556	&	0.715	&	0.534	&	0.676	&	1.188	\\
BVSP		&	0.616	&	0.581	&	1.562	&	0.601	&	0.472	&	0.932	&	0.587	&	0.455	&	1.492	&	0.615	&	0.465	&	0.939	&	0.592	&	0.666	&	0.943	&	0.550	&	0.643	&	0.917	\\
DJI			&	0.572	&	0.826	&	0.579	&	0.598	&	0.643	&	0.883	&	0.576&	0.586	&	0.970	&	0.560	&	0.661	&	0.887	&	0.581	&	0.669	&	1.230	&	0.520	&	0.690	&	0.720	\\
FCHI		&	0.617	&	0.656	&	0.969	&	0.526	&	0.621	&	1.257	&	0.579	&	0.613	&	1.087	&	0.620	&	0.579	&	1.090	&	0.553	&	0.535	&	0.894	&	0.506	&	0.633	&	1.174	\\
GDAXI		&	0.606	&	0.612	&	0.682	&	0.556	&	0.619	&	0.886	&	0.574	&	0.530	&	0.986	&	0.616	&	0.538	&	1.034	&	0.555	&	0.485	&	1.397	&	0.534	&	0.648	&	1.176	\\
GSPC		&	0.590	&	0.787	&	0.709	&	0.565	&	0.539	&	0.856	&	0.557	&	0.635	&	0.760	&	0.528	&	0.573	&	0.718	&	0.551	&	0.627	&	1.416	&	0.528	&	0.605	&	0.782	\\
GSPTSE	&	0.611	&	0.632	&	0.683	&	0.587	&	0.647	&	0.841	&	0.581	&	0.618	&	0.956	&	0.587	&	0.552	&	0.733	&	0.554	&	0.681	&	0.775	&	0.585	&	0.613	&	0.928	\\
HIS			&	0.582	&	0.823	&	0.828	&	0.562	&	0.669	&	0.639	&	0.514	&	0.730	&	0.864	&	0.592	&	0.509	&	1.083	&	0.576	&	0.749	&	0.878	&	0.557	&	0.609	&	0.805	\\
IPSA		&	0.654	&	0.969	&	0.832	&	0.584	&	0.747	&	0.984	&	0.580	&	0.519	&	1.250	&	0.582	&	0.705	&	0.938	&	0.611	&	0.677	&	1.174	&	0.601	&	0.825	&	0.801	\\
IXIC		&	0.641	&	0.707	&	0.764	&	0.585	&	0.644	&	0.941	&	0.615	&	0.702	&	0.781	&	0.563	&	0.671	&	1.425	&	0.587	&	0.680	&	1.134	&	0.591	&	0.624	&	0.901	\\
JKSE		&	0.598	&	0.848	&	1.352	&	0.539	&	0.674	&	1.335	&	0.582	&	0.725	&	0.877	&	0.560	&	0.566	&	1.802	&	0.500	&	0.907	&	0.881	&	0.570	&	0.518	&	0.769	\\
KS11		&	0.607	&	0.707	&	1.190	&	0.539	&	0.421	&	1.140	&	0.540	&	0.637	&	1.195	&	0.590	&	0.535	&	1.026	&	0.526	&	0.616	&	2.180	&	0.530	&	0.633	&	0.945	\\
MERV		&	0.651	&	0.520	&	0.927	&	0.537	&	0.625	&	1.265	&	0.537	&	0.681	&	1.163	&	0.611	&	0.647	&	0.652	&	0.540	&	0.602	&	1.135	&	0.574	&	0.534	&	0.985	\\
MXX			&	0.580	&	0.805	&	0.890	&	0.542	&	0.666	&	1.088	&	0.548	&	0.577	&	1.039	&	0.606	&	0.690	&	1.150	&	0.552	&	0.557	&	0.967	&	0.548	&	0.617	&	0.951	\\
N225		&	0.584	&	0.472	&	1.041	&	0.573	&	0.745	&	0.714	&	0.550	&	0.639	&	0.901	&	0.614	&	0.505	&	0.732	&	0.553	&	0.530	&	0.804	&	0.539	&	0.406	&	0.559	\\
NYA			&	0.593	&	0.685	&	0.466	&	0.579	&	0.648	&	0.790	&	0.550	&	0.588	&	0.615	&	0.559	&	0.691	&	0.827	&	0.526	&	0.573	&	0.954	&	0.522	&	0.583	&	0.772	\\
TWII		&	0.659	&	0.474	&	1.661	&	0.594	&	0.564	&	1.453	&	0.519	&	0.453	&	1.069	&	0.540	&	0.494	&	1.584	&	0.503	&	0.764	&	1.303	&	0.539	&	0.491	&	1.053	\\
\hline
\end{tabular}
}
\end{table}

\subsection{Comparison to bulk behavior}
The day-resolved multifractal spectra can also be compared to those for the whole time series. The motivation for such a comparison is to provide more insight on the relation between multifractality and the day-of-the-week effect. From Fig.~\ref{figcomplexity} we find that many markets (IPSA, KS11, GSPTSE, MMX) exhibit distinct multifractal properties for a particular day (e.g., Monday returns), while the whole series displays similar multifractal behavior as the bulk -- the remaining days of the week. For other markets (DJI, AXJI, N225) the overall multifractality of the series differs widely from the multifractal spectrum for each day of the week. This suggests that day-of-the-week effects result in different multifractality for these markets. We can further classify the markets into one of two multifractal behaviors: (i) bulk multifractality that only differs for one particular day of the week and (ii) day-of-the-week multifractality that is unique to every day and differs from the bulk behavior.

\subsection{Source of multifractality}
\begin{figure}[!b]
\begin{center}
\includegraphics[width=0.7\textwidth]{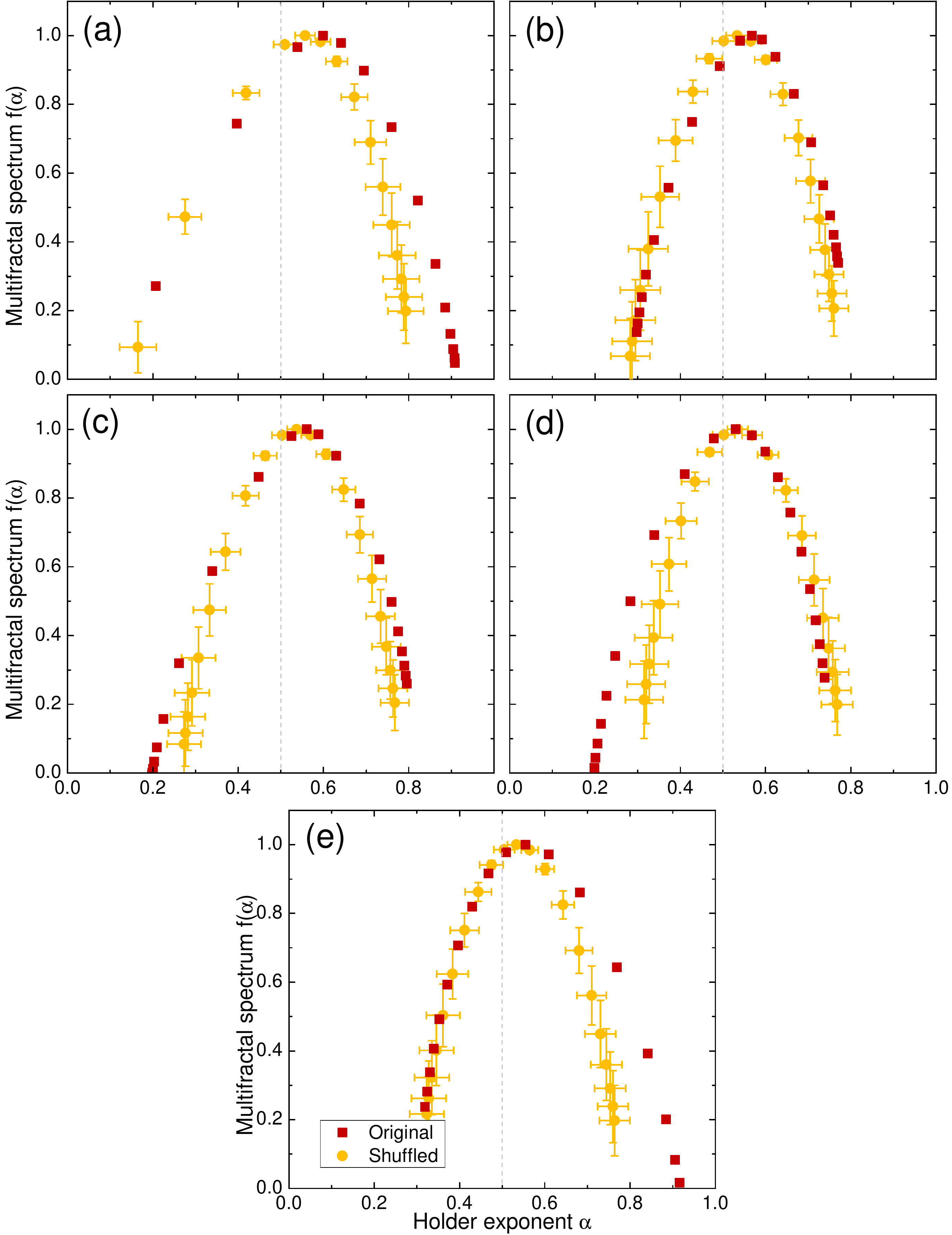}
\caption{Original and shuffled multifractal spectra $f(\alpha)$ for (a) Monday, (b) Tuesday, (c) Wednesday, (d) Thursday and (e) Friday day-resolved price returns of United States (GSPC) market.
\label{figshuffle}}
\end{center}
\end{figure}
We shuffle the time series of day-resolved returns for the four markets and then apply MF-DFA to determine the source of multifractality. The shuffling procedure performs $1000\times N$ transpositions on each series and is repeated $100$ times with different random number generator seeds, in order to obtain statistics such as the mean and standard deviation. Fig.~\ref{figshuffle} reveals that for United States (GSPC) the width of the multifractal spectrum becomes shorter during Monday and Friday returns, while the position remains the same for all of the day-resolved returns. This indicates that multifractality arises both from a broad probability density function and long-term correlations on some days, and only from probability density functions on others. Tab.~\ref{tabshuffle} lists the changes in spectra position ($\Delta\alpha_0$) and width ($\Delta W$) after shuffling day-resolved returns for GSPC, KS11, IPSA and FCHI. We find that Monday returns tend to exhibit the strongest effect from shuffling, where besides the probability density function, long-term correlations also contribute to multifractality.

\begin{table}[!t]
\caption{Differences in multifractal parameters between original and shuffled day-resolved price returns.}
\label{tabshuffle}
\centering
\scalebox{0.7}{
\begin{tabular}{lcccccccccc}
\hline
\multicolumn{1}{l}{} & \multicolumn{2}{c}{Monday} & \multicolumn{2}{c}{Tuesday} & \multicolumn{2}{c}{Wednesday} & \multicolumn{2}{c}{Thursday} & \multicolumn{2}{c}{Friday} \\
\hline
Market & $\Delta\alpha_0$ & $\Delta W$ & $\Delta\alpha_0$ & $\Delta W$ & $\Delta\alpha_0$ & $\Delta W$ & $\Delta\alpha_0$ & $\Delta W$ & $\Delta\alpha_0$ & $\Delta W$ \\ 
\hline
GSPC & 0.049 & 0.115 & 0.030 & 0.019 & 0.021 & 0.094 & 0.008 & 0.058 & 0.014 & 0.126 \\								
KS11 & 0.033 & 0.010 & 0.034 & 0.219 & 0.028 & 0.052 & 0.012 & 0.138 & 0.044 & 0.052 \\
IPSA & 0.073 & 0.200 & 0.022 & 0.119 & 0.028 & 0.069 & 0.023 & 0.064 & 0.051 & 0.098 \\
FCHI & 0.064 & 0.035 & 0.023 & 0.078 & 0.031 & 0.054 & 0.066 & 0.033 & 0.002 & 0.025 \\
\hline
\end{tabular}
}
\end{table}

\subsection{Time evolution}
\begin{figure}[!htb]
\begin{center}
\includegraphics[width=0.65\textwidth]{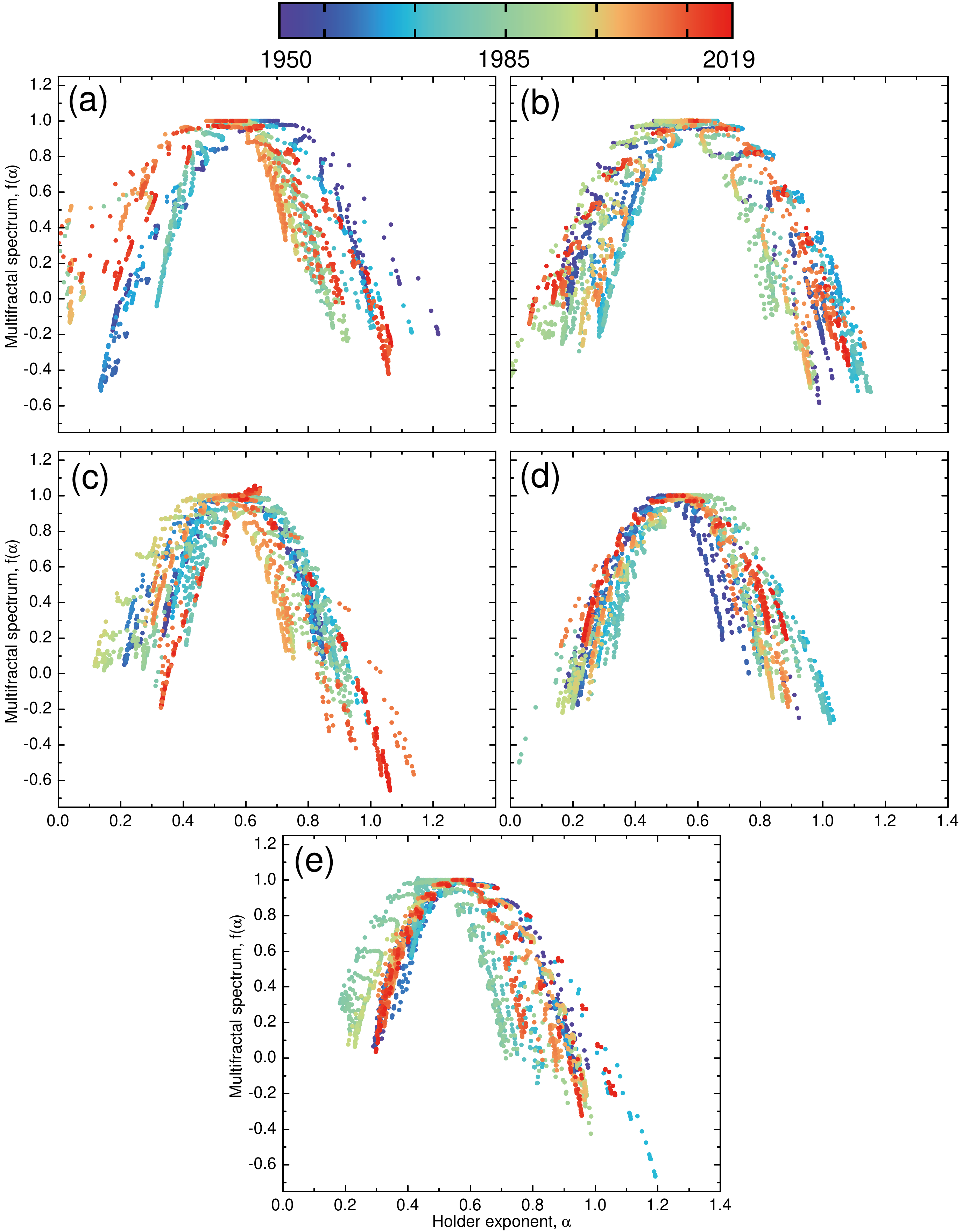}
\caption{Time evolution of the multifractal spectrum $f(\alpha)$ for (a) Monday, (b) Tuesday, (c) Wednesday, (d) Thursday and (e) Friday day-resolved price returns of United States (GSPC) market. A sliding window of 14 years and at monthly intervals is used for the period spanning 1950 to 2019.
\label{figspectrasliding}}
\end{center}
\end{figure}

For intuition on how multifractal day-of-the-week effects change over time, we can analyze time evolutions of the multifractal spectra. We consider the United States (GSPC) market since the day-of-the-week effects over time are well known~\cite{ref35}. For each day-of-the-week return we construct a sliding window of size $w=730$ days with sliding step $\Delta=5$ days, meaning that we apply the MF-DFA method on a 14-year period in monthly intervals. Fig.~\ref{figspectrasliding} illustrates time evolutions of the multifractal spectra for different day-resolved returns. We observe that the spectrum evolves differently for each day of the week. For Monday returns the spectrum shifts to the left, which means that fluctuations become less persistent over time. Other day-of-the-week returns either exhibit small movements in the multifractal spectra or move back to the same position after some time. For a more quantitative analysis we calculate the differences over time in complexity parameters, namely $\Delta\alpha_0$ and $\Delta W$, between Monday and other day-resolved returns. Fig.~\ref{figspectrasliding} reveals that the spectra position of Monday returns differ considerably from $\alpha_0$ of other day returns in the first $15$ years of the recorded period, but their differences drop to zero in subsequent years. This indicates the presence of strong day-of-the-week effects between 1950 and 1980 ($\Delta\alpha_0\rightarrow 0$ after 1965 where 1980 is already included because of the 14-year long sliding window), which is consistent with literature where it was found that day-of-the-week effects disappeared around 1980~\cite{ref35}. 
%\eject

\begin{figure}[!htb]
\begin{center}
\includegraphics[width=1\textwidth]{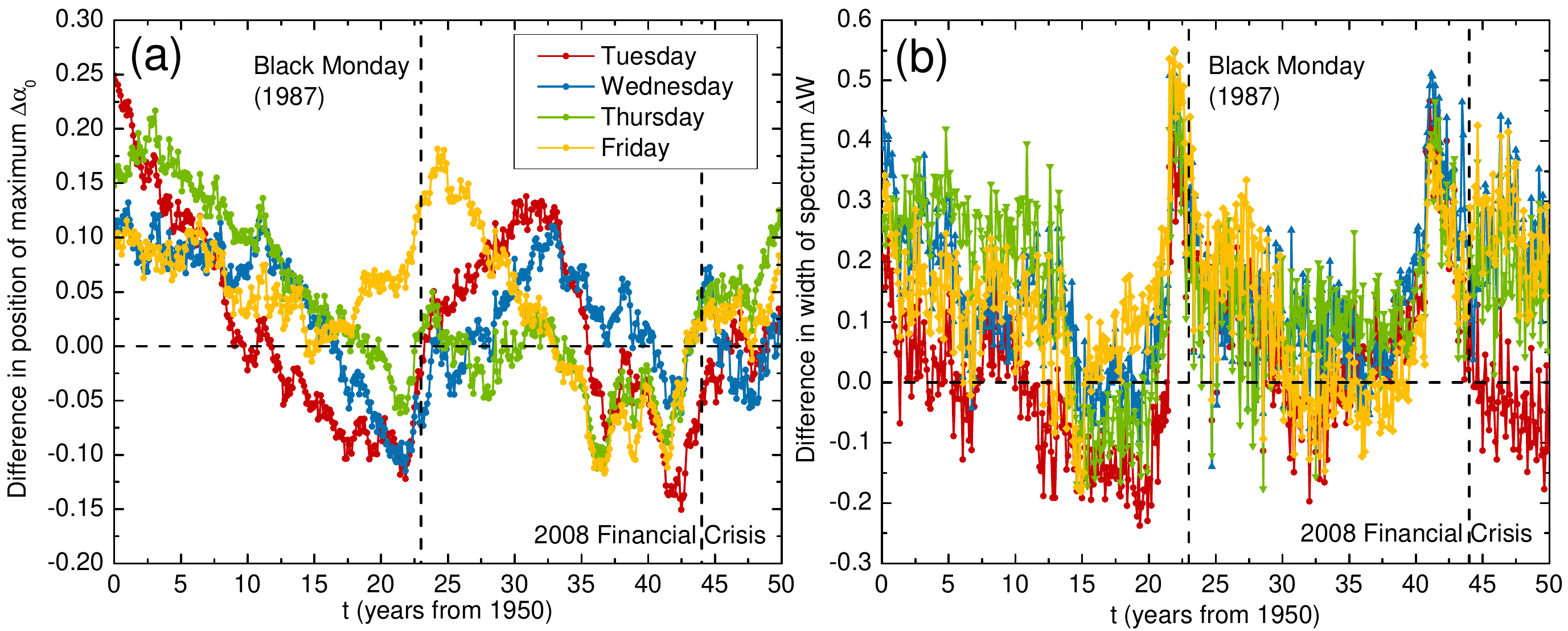}
\caption{Time evolution of differences in complexity parameters $\alpha_0$ and $W$ derived from the multifractal spectra $f(\alpha)$ between Monday and other day-resolved price returns for United States (GSPC) market. A sliding window of 14 years and at monthly intervals is used for the period spanning 1950 to 2019.
\label{figcomplexitysliding}}
\end{center}
\end{figure}

Fluctuations around $\Delta\alpha_0=0$ after 1980 can be attributed to large financial crises that affected the entire market, such as Black Monday in 1987 and the global financial crisis in 2008. Fig~\ref{figcomplexitysliding} illustrates time evolutions of differences in the spectra width $\Delta W$ between Monday and other day-resolved returns. We observe that Monday returns exhibit much wider multifractal spectra than other day returns during either of the two financial crises in 1987 and 2008. Monday returns are characterized by more complex structures and have significant day-of-the-week effects during financial crises, even after 1980 when effects from calendar anomalies should vanish. A possible explanation for this phenomena is the weekend gap in trading hours which leads to even more speculative behavior from investors during a crisis.

\section{Conclusions}\label{secconcl}
In this paper we investigate the multifractal behavior of day-of-the-week returns for market indices around the world. We apply the MF-DFA method to daily returns for each day of the week (Monday returns, Tuesday returns and so on) and calculate the multifractal spectra as well as their complexity parameters. Distinct multifractal properties are found for the different days of the week, where Monday returns typically exhibit more persistent behavior and richer multifractal structures than other day-resolved returns. We find that multifractality arises both from a broad probability density function and from long-term correlations. The time-dependent multifractal analysis on the United States (GSPC) market reveals that multifractal spectra for Monday returns shift considerably until the 1980s, and they are much wider than for other days during periods of financial crises. These results indicate that multifractality can be used as a proxy for calendar anomalies, such as the day-of-the-week effect often considered in financial literature. Future studies should be directed at further investigating the multifractal dynamics and day-of-the-week effects for other financial markets, and extending the current analysis to other calendar anomalies.

\section{Acknowledgments}\label{secack}
T.S. acknowledges support of Brazilian agency CNPq, grant No 304497/2019-3.

%\eject

%% References


\bibliographystyle{elsarticle-num}
\bibliography{dayofweekmfdfa}

\end{document}